\documentclass[sigconf]{acmart}
\usepackage{cleveref}

\AtBeginDocument{%
  \providecommand\BibTeX{{%
    \normalfont B\kern-0.5em{\scshape i\kern-0.25em b}\kern-0.8em\TeX}}}

\copyrightyear{2020}
\acmYear{2020}
\setcopyright{acmcopyright}
\acmConference[ICPE '20 Companion]{ACM/SPEC International Conference on
Performance Engineering Companion}{April 20--24, 2020}{Edmonton, AB, Canada}
\acmBooktitle{ACM/SPEC International Conference on Performance
Engineering Companion (ICPE '20 Companion), April 20--24, 2020,
Edmonton, AB, Canada}
\acmPrice{15.00}
\acmDOI{10.1145/3375555.3384935}
\acmISBN{978-1-4503-7109-4/20/04}

\begin{document}
\fancyhead{} 
\title{ Extended Abstract of Performance Analysis and Prediction of Model Transformation}








\author{Vijayshree Vijayshree}
\affiliation{\institution{University of Stuttgart}}
\email{vijayshree.vijayshree@iste.uni-stuttgart.de}
\author{Markus Frank}
\affiliation{\institution{University of Stuttgart}}
\email{markus.frank@iste.uni-stuttgart.de}
\author{Steffen Becker}
\affiliation{\institution{University of Stuttgart}}
\email{steffen.becker@iste.uni-stuttgart.de}
\renewcommand{\shortauthors}{Trovato and Tobin, et al.}

\maketitle

\section{Introduction}
Nowadays model transformation techniques are a popular technique within Model-Driven Engineering (MDE). 
Model transformations enable the generation of new models, realizing changes on individual models, and the synchronization between models. In the model transformations there exist various transforming languages, our focus is on Query View Transformation Operational (QVTo).  

However, a system being developed with model transformations can grow in size and can become complex. For example, in an automotive domain, an AUTOSAR model of a large electronic control unit (ECU) for modern cars has over 170.000 model elements. In such cases, the execution of badly performing model transformations can take up to hours. Hence, performance is an important quality attribute of model transformations, e.g., execution time, memory usage.
The existing technique\cite{burmester2005worst} to identify and visualize the root cause of low performing transformation rules are not addressed properly, meanwhile, performance optimization is limited only to the transformation engine. Hence, model transformation rules are considered to be fixed and unchangeable. Unfortunately, transformation engineers have no insights about  how long the transformation takes place and also to predict  the performance.

To the best of our knowledge, there exists no research which makes analyzes of the transformation rules and presents to the transformation engineers. 

Therefore it is our vision to provide an approach towards performance engineering of model transformation. Hence, this approach will enable us to systematically \textit{monitor and visualize causes for performance issues as well as predict the performance of model transformations}.

We will demonstrate and identify the root cause of low performing QVTo rules with the help of three different phases. Firstly, in the monitoring framework, we measure the execution time of each QVTo transformation rules. Secondly, the profiler will graphically visualize the results. Finally, we will demonstrate to predict the overall execution time of QVTo transformations with the data stored in the database. 






\section{Related Work}
We extensively focus on state of the art closely related to optimization of the performance of model transformation engine.

A class of approaches \cite{burmester2005worst} analyzes model transformations and compute worst-case execution time upon optimal search order of the story pattern elements.  Also, another approache
\cite{tichy2013detecting} of the Henshin interpreter, considers models to execute the model transformations and thus address the bad smells which affects the performance of model transformation.

Amstel \cite{van2010metrics} investigates the factors such as size and complexity of the input models affecting the performance of model transformations. The author has compared different languages and systematically analyzed the influence of extracted metric value on performance of model transformations.     

Piers \cite{piers2010atl} provides a way to detect performance issues on a ATL transformation by a detailed analysis of the execution. An execution profile stores information about the execution time, the memory used, etc of the model transformation.

Becker \cite{becker2008model} made analysis performance  and prediction, by generating prototypes from models, which in turn generate code skeleton or require detailed models for the prototype.

Groner \cite{groner2018towards} provides possible visualization and refactoring methods to improve the performance of model transformation in a declarative way.

Hence, these approaches show a significant performance improvement by refactoring the engine of the model transformations. On the other hand lacks the measurements and refactoring of transformation  rules. Hence, in this paper, we provide an approach for supporting transformation engineers in identifying the root causes. Complementary helping engineers improve the transformation rules by themselves,  which leads to the performance gain.

\section{Proposed Approach} 
\begin{figure}
\centering
\includegraphics[width=0.5\textwidth]{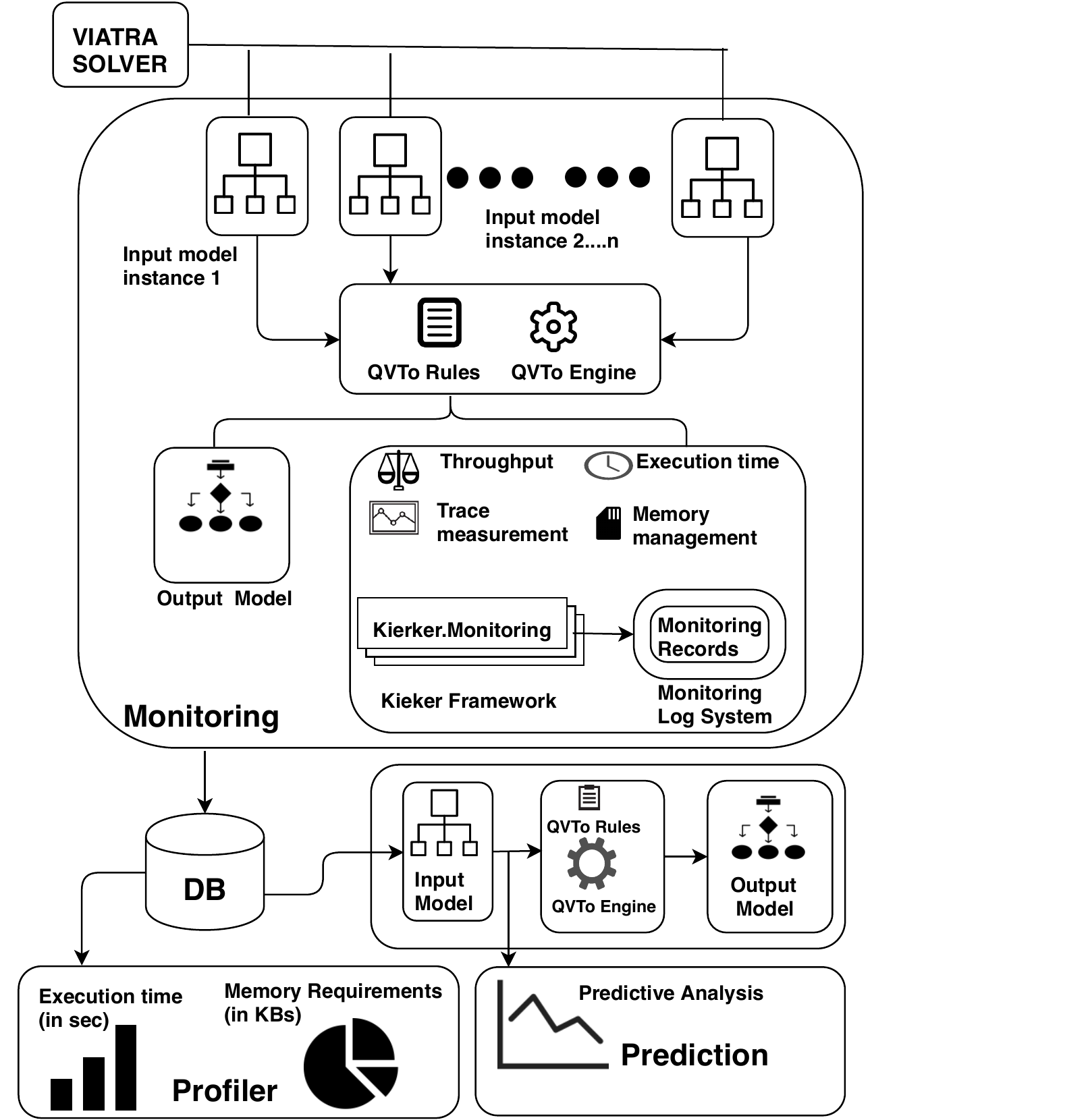}
\caption{ Proposed Approach}
\label{fig:GSPN}
\end{figure}

In the proposed approach we will contribute towards improving the performance of model transformation in an imperative way. 
Fig. \ref{fig:GSPN} explains the three phases of our approach namely \textbf{Monitoring}, \textbf{Profiling}, and \textbf{Prediction}.
To generate a large test set of input instance models we use VIATRA solver \cite{semerath2018graph}.
 Thus, the generated instances are then transformed into the respected output model with the help of QVTo rules and run by the QVTo engine. During the execution we using the Kieker monitoring framework \cite{van2012kieker} to gather all the necessary operational profiles \cite{van2010metrics} and place them in the database. The data from the database are visualized to identify the performance bottlenecks. In turn, the data from the database is used to predict the performance before the actual transformations.

\textbf{Monitoring WP1:} To analyze the performance of transformations, engineers need to learn about the operational profile  \cite{van2010metrics} of the model transformations. The operational profile includes resource demands like execution times, rule evaluations, time spent in a model I/O. To measure the operational profile, we will extend the QVTo engine by injecting pointcuts of aspect-oriented programming \cite{kiczales1997aspect} to the rules, whenever the engine executes these rules, Kieker monitoring API \cite{van2012kieker} is executed, to fetch operational profiles and measure them. Once the operational profile is measured they are stored in the database for further analysis and prediction.

\textbf{Profiling WP2:} The \textbf{WP1} will provide the raw and too detailed data about the operational profile of the model transformations. The raw data contains the execution time of each transformation rule, overall execution time. This detailed data will not directly help the transformation engineer in understanding where exactly lies the performance issues. Hence, this raw-data, in turn, needs to be visualized to support the engineer in identifying the root cause of badly performing transformation rules. Therefore, we are designing the profiler which presents the analyzed raw-data to the transformation engineer. A performance decline can be the result of changes in the model transformation or by changes in the meta-model or the operational profile. With the help of the developed profiler, we can easily identify and rank a list of possible causes for the performance decline by using monitoring data.

\textbf{Prediction WP3:} To support the engineer in predicting the performance, we need to develop a prediction framework. The developed framework will help to predict the performance change of model transformations. 
To predict the performance of a model without having a prior reference model or historic operational profile data of previously transformed model is always a difficult job and performance prediction of a model may not be accurate. To generate the different instance reference model we need to scale the input model either automatically or manually. However, scaling manually is always error-prone and tedious job while we need to be very specific about the dependencies of the scaling elements. Hence, to overcome such a problem we are reusing the existing VIATRA \cite{semerath2018graph} tool to automatically generate the instances of the input model, each instance is different. Then each instance model is transformed to obtain an output model. Subsequently, the operational profile (e.g., execution time and memory usage) of each instance model is obtained and thus, data are stored in the database. Eventually, the complete setup of generating instances, transforming the model and measurements of the operational profile is run in a continuous integration environment at a defined interval of time, which in turn serves as the reference data and thus, helps in performance prediction of transformations.

\section{Conclusion}
In this paper, we demonstrated an approach to identify the root cause of low performing QVTo rules. 
We presented the three phases of our approach namely monitoring, profiling, and prediction. In the monitoring phase, we will systematically monitor all the operational profile with the help of aspect- oriented pointcuts and kieker framework. In the profiling phase,  we visualize the monitored operational profile of the monitoring phase and support the transformation engineer to identify root cause of badly performing transformation rules. With the use of VIATRA solver, we will automatically generate instances of input model and perform model transformations to measure the operational profiles and store them in database. This particular monitored data will be used for the prediction purposes.
\section{ACKNOWLEDGEMENT}
This work was funded by the Deutsche Forschungsgemeinschaft
(DFG, German Research Foundation)- BE 4796-3-1.
\bibliographystyle{ACM-Reference-Format}
\bibliography{sample-base}

\appendix









\end{document}